\documentclass[cameraready]{Interspeech}
\title{Personalized Keyword Spotting for User-Defined Keywords Leveraging Text-Independent Speaker Verification}

\author[affiliation={1}, orcid=0009-0004-6936-2576]{Ming-Hsiang}{Hu}
\author[affiliation={1},orcid=0009-0003-1417-9638]{Kuan-Tang}{Huang}
\author[affiliation={2}, orcid=0009-0000-1392-0058]{Chien-Chun}{Wang}
\author[affiliation={3}, orcid=0000-0001-7044-9434]{Hung-Shin}{Lee}
\renewcommand{\authorsep}{\\}
\author[affiliation={1}, orcid=0000-0003-0693-8932]{Berlin}{Chen}
\address{
  $^{1}$Dept. Computer Science and Information Engineering, National Taiwan Normal University, Taiwan \\
  $^{2}$E.SUN Financial Holding Co., Ltd., Taiwan \\
  $^{3}$United Link Co., Ltd., Taiwan
}

\email{}


\newcommand{\component}[1]{\vspace{3pt}\noindent\textbf{#1.} }
\usepackage{comment}
\usepackage{amsmath}
\usepackage{graphicx}
\usepackage{cite}
\usepackage{multirow}
\usepackage{booktabs}
\usepackage{amssymb}
\usepackage{enumitem}
\usepackage{url}
\usepackage{hyperref}
\usepackage{xcolor}
\usepackage[utf8]{inputenc}
\usepackage[table]{xcolor} % 用於顏色處理
\usepackage{tabularx}

\definecolor{tablegreen}{HTML}{D9EAD3}
\definecolor{tablered}{HTML}{F4CCCC}
\definecolor{tablegray}{HTML}{EFEFEF}

\begin{document}

\maketitle

\begin{abstract}
  User-defined keyword spotting (UD-KWS) enables zero-shot wake-word detection from text, but existing systems learn speaker-invariant representations that cannot reject impostors uttering the correct keyword. We address this dual zero-shot setting---unseen keywords and unseen speakers---with ZP-KWS, a lightweight framework combining a phoneme-supervised audio encoder with a GE2E-pretrained compact speaker encoder (about 0.9M parameters). Multiplicative late fusion at inference grants each branch independent veto power, supporting modes from conventional detection to strict speaker-gated activation without retraining. On LibriPhrase, Google Speech Commands, and Qualcomm datasets, ZP-KWS reduces target-only FRR at 1\% FAR by up to 60\% relative to the strongest baseline while maintaining competitive keyword detection, all within a 1.55M parameter budget for edge deployment.
\end{abstract}

\section{Introduction}
Voice user interfaces require activation mechanisms that are efficient, secure, and personalized.
Conventional KWS has long focused on compact on-device models \cite{chen2014}.
User-defined keyword spotting (UD-KWS) extends this line by enabling zero-shot wake-word detection from arbitrary text inputs, eliminating retraining for new keywords~\cite{jung2023,navon2024,nishu2024,xi2024,jung2025,ai2026,jung2026}.
Cross-modal alignment methods such as CMCD~\cite{shin2022} and PhonMatchNet~\cite{kreuk2023} improve UD-KWS by matching text-derived phoneme sequences to audio representations, while recent studies further explore multi-modal fusion~\cite{ai2024} and dual-data scaling~\cite{ai2026}.

However, learning text-invariant representations often suppresses speaker cues.
As a result, these systems cannot reject impostors---for example, bystanders or playback audio---who utter the correct keyword.
In edge deployment, such false activations degrade user experience and waste power.
PK-MTL~\cite{yang2022} partially addresses this issue by jointly training a shared encoder for keyword spotting and text-dependent speaker verification (TD-SV), then combining keyword and speaker scores with a task-specific scoring function.
This TD-SV formulation is effective when enrollment and test utterances share the same pre-defined keyword, because TD-SV benefits from phonetic alignment.
Yet the text-dependent assumption constrains PK-MTL to fixed keywords: when the keyword changes, both the KWS classifier and speaker enrollment must be updated, which breaks the zero-shot flexibility expected in UD-KWS.

Extending personalization to UD-KWS therefore requires text-independent SV (TI-SV), where enrollment and query utterances can have different phonetic content.
This setting creates a dual zero-shot problem with unseen keywords and unseen speakers.
Although TI-SV is mature for long-form speech\cite{snyder2018, desplanques2020,liu2022,wang2023,chen2024,miara2024}, discriminability drops substantially on sub-second utterances.
In addition, typical TI-SV models ($>$10\,M parameters) remain too heavy for edge devices.

To address these challenges, we propose \textbf{ZP-KWS}\footnote{The source code is available at: \url{https://github.com/Padawan101/ZP-KWS}}, a lightweight framework for \textbf{Z}ero-shot \textbf{P}ersonalized \textbf{KWS}.
Our contributions are:
\begin{enumerate}[noitemsep,leftmargin=*]
  \item \textbf{Sub-second TI-SV via GE2E Pre-training:}
    We show that optimizing an ultra-compact encoder (EfficientTDNN-Small~\cite{wang2022}, $\sim$0.9\,M parameters) with the GE2E loss~\cite{wan2018} stabilizes high-variance sub-second embeddings, yielding a 62\% relative EER reduction over standard pooling.
  \item \textbf{Phoneme-Supervised Audio Encoder:} Frame-level phoneme supervision strengthens phonetic structure in the trainable audio stream and improves discrimination on hard minimal pairs.
  \item \textbf{Modular Multiplicative Fusion:} A multiplicative late-fusion strategy enforces a strict AND condition: both keyword content and speaker identity must be independently validated.
    This multiplicative-fusion design enables dynamic switching between conventional and target-only modes without retraining and reduces TO-KWS FRR at 1\% FAR by up to 60\% relative.
\end{enumerate}

\begin{figure*}[ht]
  \centering
  \includegraphics[width=0.9\linewidth]{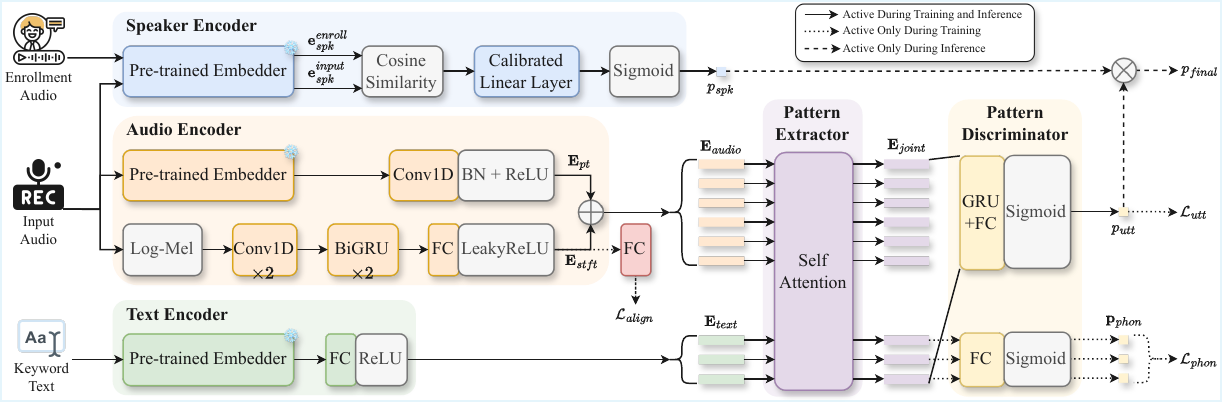}
  \vspace{-5pt}
  \caption{
    Architecture of ZP-KWS. Auxiliary phoneme supervision ($\mathcal{L}_{align}$, dotted) is training-only; multiplicative late fusion ($p_{final}$, dashed) is inference-only.
  }
  \label{fig:main}
  \vspace{-15pt}
\end{figure*}

\section{Proposed Method}
\subsection{Framework Overview}

To address the dual zero-shot challenge while minimizing task interference, ZP-KWS uses two functionally separated branches: a TI-SV branch for speaker identity and a phoneme-supervised branch for keyword content.
The core design choice is inference-time late fusion (dashed path in Figure \ref{fig:main}).
Instead of additive score fusion, where a high keyword score can compensate for a low speaker score, we enforce a strict AND condition so that both constraints must be satisfied.
The final decision score $p_{final}$ multiplicatively combines the semantic $p_{utt}$ and acoustic identity $p_{spk}$ probabilities:
\vspace{-5pt}
\begin{equation}
  p_{final} = p_{utt} \cdot p_{spk},
  \label{eq:fusion}
\end{equation}
where $p_{utt}, p_{spk} \in [0,1]$.
This inference-time fusion design gives each branch veto power: activation occurs only when both keyword content and speaker identity are validated.
By adjusting thresholds on these probabilities, we support three operating modes without retraining: conventional (C-KWS, $p_{spk} \equiv 1$), target-biased (TB-KWS), and target-only (TO-KWS).

\subsection{Model Architecture}

\component{Speaker Encoder}
To meet edge constraints, we adopt EfficientTDNN-Small \cite{wang2022} ($\sim$0.9\,M parameters) to extract 192-dimensional speaker embeddings.
We freeze this encoder during KWS training to prevent semantic-task gradients from distorting the speaker embedding space.
At evaluation, we enroll each target speaker with a single utterance whose keyword may differ from the query, matching a practical text-independent setup.
Given embeddings for the enrollment ($\mathbf{e}^{enroll}_{spk}$) and query ($\mathbf{e}^{input}_{spk}$), the speaker match probability $p_{spk}$ is computed via a logistic mapping:
\begin{equation}
  p_{spk} = \sigma\left(w_{spk} \cdot \cos(\mathbf{e}^{input}_{spk},\, \mathbf{e}^{enroll}_{spk}) + b_{spk}\right),
  \label{eq:pspk}
\end{equation}
where $\sigma(\cdot)$ is the sigmoid function, and $w_{spk}, b_{spk}$ are fixed hyperparameters that map raw cosine similarity to the $[0,1]$ probability range aligned with $p_{utt}$.
Because statistical pooling degrades on sub-second queries with limited temporal context, we train the speaker encoder in two stages.
First, EfficientTDNN-Small is pre-trained on VoxCeleb2~\cite{chung2018} to acquire speaker discrimination across diverse acoustic conditions.
We then fine-tune it with GE2E loss~\cite{wan2018} on the 460\,h LibriPhrase training set, where utterances are predominantly sub-second keyword segments.
For a batch of N speakers with M utterances per speaker, GE2E pulls each embedding toward its own speaker centroid and pushes it away from other in-batch centroids, which stabilizes high-variance short-utterance embeddings.

\component{Audio Encoder}
The audio encoder extracts frame-level representations $\mathbf{E}_{audio} \in \mathbb{R}^{T_a \times 128}$ ($T_a$ denotes acoustic frames) through two parallel streams.
A frozen pre-trained embedder~\cite{lin2020} computes 96-dimensional features every 80\,ms, upsampled to the 20\,ms frame rate via transposed convolution and linear projection following~\cite{kreuk2023}, yielding $\mathbf{E}_{pt}$.
Concurrently, a trainable stream processes 40-dimensional log-Mel features through two Conv1D layers (128 and 256 channels, kernel size 5, each followed by batch normalization and ReLU) and a two-layer bidirectional GRU (128 hidden units per direction), followed by a fully-connected projection to 128 dimensions, yielding $\mathbf{E}_{stft}$.
We fuse the streams as $\mathbf{E}_{audio} = \mathbf{E}_{pt} + \mathrm{LayerNorm}(\mathbf{E}_{stft})$, where LayerNorm\cite{ba2016} controls the scale of $\mathbf{E}_{stft}$ so it does not dominate the frozen $\mathbf{E}_{pt}$.
To further strengthen phonetic discrimination, $\mathbf{E}_{stft}$ is projected through a fully-connected (FC) layer to predict phoneme posteriors, supervised by $\mathcal{L}_{align}$ (Section~\ref{sec:align}).

\component{Text Encoder}
We convert each keyword to phonemes with a grapheme-to-phoneme (G2P) model\footnote{\url{https://github.com/Kyubyong/g2p}}, then project it to semantic embeddings $\mathbf{E}_{text} \in \mathbb{R}^{T_t \times 128}$ ($T_t$ is sequence length).

\component{Pattern Extractor}
Following~\cite{kreuk2023}, the concatenated sequence $\mathbf{E}_{concat} = [\mathbf{E}_{audio};\, \mathbf{E}_{text}]$ is processed by a single-head scaled dot-product self-attention with a causal mask, producing $\mathbf{E}_{joint}$.

\component{Pattern Discriminator}
The discriminator decodes $\mathbf{E}_{joint}$ into two probabilities that operate at different granularities. We compute utterance-level match probability $p_{utt} \in [0,1]$ by passing $\mathbf{E}_{joint}$ through a single-layer GRU (128 hidden units), followed by an FC layer and sigmoid.
We compute phoneme-level match sequence $\mathbf{p}_{phon} \in \mathbb{R}^{T_t \times 1}$ position-wise on the text-aligned part of $\mathbf{E}_{joint}$ with a separate FC layer and sigmoid.

\begin{table*}[t]
  \small
  \centering
  \caption{Performance comparison evaluation of the proposed ZP-KWS against baselines across three operational modes. Metrics are ordered by practical significance for edge deployment. Best results in each metric are highlighted in \textbf{bold}.}
  \vspace{-5pt}
  \label{tab:main_results}
  \resizebox{1.0\textwidth}{!}{
    \begin{tabular}{l l | ccc | ccc | ccc}
      \toprule
      \multirow{2}{*}{\bf Dataset} & \multirow{2}{*}{\bf Method}
      & \multicolumn{3}{c|}{\bf C-KWS (\%) $\downarrow$}
      & \multicolumn{3}{c|}{\bf TB-KWS (\%) $\downarrow$}
      & \multicolumn{3}{c}{\bf TO-KWS (\%) $\downarrow$} \\
      \cmidrule(lr){3-5} \cmidrule(lr){6-8} \cmidrule(lr){9-11}
      & & FRR@1\% & EER & FRR@10\%
      & FRR@1\% & EER & FRR@10\%
      & FRR@1\% & EER & FRR@10\% \\

      \midrule
      \multirow{6}{*}{\bf \shortstack[l]{LibriPhrase\\Easy}}
      & PhonMatchNet \cite{kreuk2023}
      & 10.48 & 3.34 & 0.90
      & 10.54 & 3.36 & 0.88
      & 97.00 & 25.16 & 70.24 \\
      & PK-MTL \cite{yang2022}
      & 12.48 & 3.46 & 0.78
      & 11.43 & 3.35 & 0.71
      & 72.79 & 17.74 & 28.94 \\
      \cmidrule(lr){2-11}
      & \textbf{ZP-KWS (Ours)}
      & 6.72 & \textbf{2.38} & \textbf{0.43}
      & 6.57 & \textbf{2.36} & \textbf{0.38}
      & 29.47 & 8.16 & 6.93 \\
      & \quad w/o GE2E Pre-training
      & 5.83 & 2.45 & 0.43
      & 7.78 & 2.69 & 0.83
      & 73.42 & 18.02 & 32.49 \\
      & \quad w/o Calibrated Linear Layer
      & \textbf{5.71} & 2.46 & 0.59
      & \textbf{5.62} & 2.42 & 0.57
      & 32.39 & 15.24 & 16.78 \\
      & \quad w/o Phoneme Supervision
      & 8.15 & 2.88 & 0.62
      & 9.27 & 2.69 & 0.54
      & \textbf{28.89} & \textbf{7.92} & \textbf{6.57} \\

      \midrule
      \midrule
      \multirow{6}{*}{\bf \shortstack[l]{LibriPhrase\\Hard}}
      & PhonMatchNet \cite{kreuk2023}
      & 87.70 & 20.48 & 41.48
      & 87.68 & 20.45 & 41.33
      & 97.48 & 31.86 & 74.94 \\
      & PK-MTL \cite{yang2022}
      & 89.04 & 21.31 & 45.47
      & 90.32 & 19.38 & 40.71
      & 87.72 & 24.04 & 46.30 \\
      \cmidrule(lr){2-11}
      & \textbf{ZP-KWS (Ours)}
      & 83.11 & 17.48 & 31.69
      & \textbf{83.87} & \textbf{14.28} & 24.32
      & \textbf{78.20} & \textbf{13.87} & \textbf{19.62} \\
      & \quad w/o GE2E Pre-training
      & 84.29 & 16.89 & 29.83
      & 91.33 & 18.50 & 41.25
      & 88.88 & 24.34 & 48.95 \\
      & \quad w/o Calibrated Linear Layer
      & \textbf{80.59} & \textbf{16.72} & \textbf{29.15}
      & 86.11 & 15.85 & \textbf{19.61}
      & 79.62 & 18.92 & 23.00 \\
      & \quad w/o Phoneme Supervision
      & 85.66 & 18.19 & 34.17
      & 88.67 & 15.06 & 29.63
      & 83.80 & 13.91 & 21.02 \\

      \midrule
      \midrule
      \multirow{6}{*}{\bf Qualcomm}
      & PhonMatchNet \cite{kreuk2023}
      & 39.18 & 8.67 & 7.41
      & 38.53 & 9.04 & 8.08
      & 93.12 & 16.06 & 35.66 \\
      & PK-MTL \cite{yang2022}
      & 24.57 & 8.88 & 8.20
      & 23.04 & 8.55 & 7.65
      & 55.71 & 13.73 & 16.34 \\
      \cmidrule(lr){2-11}
      & \textbf{ZP-KWS (Ours)}
      & \textbf{23.19} & \textbf{6.88} & \textbf{4.37}
      & \textbf{17.97} & \textbf{5.67} & \textbf{2.52}
      & \textbf{33.12} & \textbf{8.81} & \textbf{7.94} \\
      & \quad w/o GE2E Pre-training
      & 40.47 & 9.38 & 8.58
      & 69.77 & 8.29 & 5.91
      & 67.44 & 9.65 & 9.54 \\
      & \quad w/o Calibrated Linear Layer
      & 29.04 & 8.30 & 6.82
      & 24.71 & 8.03 & 6.82
      & 35.96 & 15.00 & 23.10 \\
      & \quad w/o Phoneme Supervision
      & 41.30 & 10.81 & 11.80
      & 41.98 & 8.77 & 8.12
      & 43.64 & 10.51 & 11.02 \\

      \midrule
      \midrule
      \multirow{6}{*}{\bf \shortstack[l]{Google\\Speech\\Commands}}
      & PhonMatchNet \cite{kreuk2023}
      & 46.20 & \textbf{10.07} & 10.15
      & 46.48 & 10.24 & 10.43
      & 82.61 & 12.27 & 15.57 \\
      & PK-MTL \cite{yang2022}
      & 48.29 & 11.21 & 12.57
      & 51.51 & 10.91 & 11.90
      & 65.66 & 13.14 & 18.61 \\
      \cmidrule(lr){2-11}
      & \textbf{ZP-KWS (Ours)}
      & 35.96 & 10.74 & 11.44
      & 44.46 & \textbf{9.46} & \textbf{9.02}
      & \textbf{49.76} & 11.05 & 12.24 \\
      & \quad w/o GE2E Pre-training
      & 35.34 & 11.38 & 12.33
      & 65.93 & 11.18 & 12.30
      & 71.24 & 13.06 & 17.52 \\
      & \quad w/o Calibrated Linear Layer
      & 46.95 & 11.93 & 13.09
      & \textbf{44.33} & 11.80 & 12.84
      & 51.57 & 13.36 & 16.69 \\
      & \quad w/o Phoneme Supervision
      & \textbf{28.05} & 10.08 & \textbf{10.12}
      & 52.36 & 9.53 & 9.23
      & 55.14 & \textbf{10.66} & \textbf{11.35} \\

      \bottomrule
    \end{tabular}
  }
  \vspace{-14pt}
\end{table*}

\subsection{Frame-Level Phoneme Supervision}
\label{sec:align}

Downstream objectives ($\mathcal{L}_{utt}, \mathcal{L}_{phon}$) supervise keyword matching but only indirectly constrain frame-level phonetic structure in the audio encoder.
To provide explicit phonetic guidance, we add a training-only auxiliary classification head on $\mathbf{E}_{stft}$ (Figure~\ref{fig:main}). This auxiliary branch is removed at inference and therefore introduces no runtime overhead.

\component{Label Generation}
We run MFA \cite{mcauliffe2017} on the LibriPhrase training set to obtain 20\,ms frame-level targets.
The auxiliary head projects $\mathbf{E}_{stft}$ to $D=42$ dimensions.
We apply this supervision only to $\mathbf{E}_{stft}$ so that the trainable stream learns phonetic discrimination directly, rather than relying on linguistic priors already encoded in the frozen $\mathbf{E}_{pt}$.

\component{Alignment Loss Formulation}
The objective is a label-smoothed cross-entropy loss:
\vspace{-5pt}
\begin{equation}
  \mathcal{L}_{align} = -\frac{1}{|\mathcal{V}|} \sum_{t \in \mathcal{V}} \sum_{d=1}^{D} \tilde{y}_{t,d} \log \hat{y}_{t,d},
  \label{eq:ldn}
\end{equation}
where $\mathcal{V}$ denotes valid frame indices within a batch, $\hat{y}_{t,d}$ is the softmax prediction, and $\tilde{y}_{t,d}$ is the smoothed target distribution that accounts for boundary uncertainty:
\vspace{-5pt}
\begin{equation}
  \tilde{y}_{t,d} = (1 - \varepsilon) y_{t,d} + \frac{\varepsilon}{D},
  \label{eq:smooth}
\end{equation}
where $y_{t,d} \in \{0, 1\}$ is the one-hot ground truth.
The smoothing parameter $\varepsilon$\cite{szegedy2016} models temporal ambiguity near forced-alignment boundaries and reduces over-confident predictions on transition frames.

\subsection{Training Objectives}

We optimize the framework end-to-end with an unweighted multi-objective loss:
\begin{equation}
  \mathcal{L}_{total} = \mathcal{L}_{utt} + \mathcal{L}_{phon} + \mathcal{L}_{align}.
  \label{eq:loss}
\end{equation}
$\mathcal{L}_{utt}$ and $\mathcal{L}_{phon}$ are BCE losses on utterance-level $p_{utt}$ and phoneme-level $\mathbf{p}_{phon}$ predictions, respectively, and jointly optimize keyword matching at coarse and fine granularity.
$\mathcal{L}_{align}$ (Eq.~\ref{eq:ldn}) adds explicit frame-level phonetic supervision for the trainable audio stream.
To prevent cross-branch gradient interference, we restrict gradients from $\mathcal{L}_{align}$ to the trainable stream only, so phoneme-supervision updates do not perturb the frozen pre-trained stream.

\section{Experiments}
\subsection{Experimental Setup}

\component{Datasets}
We evaluated ZP-KWS on LibriPhrase \cite{shin2022} (Easy and Hard splits) as the in-domain benchmark.
To test out-of-domain generalization, we additionally used Google Speech Commands (GSC) \cite{warden2018} and Qualcomm Keyword Speech \cite{kim2019}, which introduce different acoustic conditions and vocabularies.
In all evaluations, both target keywords and target speakers were unseen during training, enforcing a strict dual zero-shot setting.

\component{Evaluation Protocol}
Following PK-MTL~\cite{yang2022}, each trial was defined by two binary attributes: keyword match and speaker match.
This yields four pair types (Table~\ref{tab:task}): target-speaker/target-keyword (\textbf{ts-tk}), non-target-speaker/target-keyword (\textbf{nts-tk}), target-speaker/non-target-keyword (\textbf{ts-ntk}), and non-target-speaker/non-target-keyword (\textbf{nts-ntk}).
Using these four pair types, we defined three operating modes with increasing strictness:
\textbf{C-KWS} (conventional) accepts any speaker if the keyword is correct;
\textbf{TB-KWS} (target-biased) treats \textbf{nts-tk} as neutral (neither positive nor negative);
\textbf{TO-KWS} (target-only) accepts only \textbf{ts-tk} and rejects all other cases, including impostors who speak the correct keyword.
For all datasets, we maintained a balanced 1:1 target-to-non-target speaker ratio when forming evaluation pairs.

\begin{table}[t]
  \centering
  \footnotesize
  \caption{Categorization of evaluation pairs and the three operational modes.}
  \label{tab:task} % 增加 label 以便在文中引用
  \vspace{-8pt}
  \renewcommand{\arraystretch}{1.3}
  \begin{tabularx}{\linewidth}{|>{\centering\arraybackslash}X|>{\centering\arraybackslash}X|>{\centering\arraybackslash}X|>{\centering\arraybackslash}X|>{\centering\arraybackslash}X|}
    \hline
    & \textbf{ts-tk} & \textbf{nts-tk} & \textbf{ts-ntk} & \textbf{nts-ntk} \\ \hline
    \textbf{C-KWS} & \cellcolor{tablegreen} O & \cellcolor{tablegreen} O & \cellcolor{tablered} X & \cellcolor{tablered} X \\ \hline
    \textbf{TB-KWS} & \cellcolor{tablegreen} O & \cellcolor{tablegray} - & \cellcolor{tablered} X & \cellcolor{tablered} X \\ \hline
    \textbf{TO-KWS} & \cellcolor{tablegreen} O & \cellcolor{tablered} X & \cellcolor{tablered} X & \cellcolor{tablered} X \\ \hline
  \end{tabularx}
  \vspace{-18pt}
\end{table}

\component{Baseline Configuration}
PhonMatchNet~\cite{kreuk2023} served as the UD-KWS backbone for all systems, ensuring a fair comparison with a shared feature extractor.
For PK-MTL~\cite{yang2022}, originally designed for fixed-vocabulary KWS, we replaced its classifier-based keyword backbone with PhonMatchNet so it could operate in a zero-shot keyword setting.
Speaker scores were produced by PK-MTL's Score Combination Module (SCM), and the linear fusion coefficient $\alpha$ was tuned on the validation set.
This configuration yielded the strongest PK-MTL variant in our text-independent evaluation, making it a stronger baseline than the original PK-MTL implementation.

\component{Implementation Details}
We used 40-dimensional log-Mel spectrograms (25\,ms window, 10\,ms hop).
All trainable components in ZP-KWS were optimized end-to-end using AdamW (learning rate $10^{-4}$, batch size 2048).
All experiments were run on a single NVIDIA RTX 5090 GPU.
The loss terms in Eq.~\ref{eq:loss} were equally weighted, and label smoothing was set to $\varepsilon=0.1$.
The logistic mapping hyperparameters in Eq.~\ref{eq:pspk} were set to $w_{spk}=10$ and $b_{spk}=-5$, corresponding to a cosine-similarity decision boundary of $0.5$ based on the training-set score distribution.
These values were selected by grid search on the LibriPhrase training split only, aligning $p_{spk}=0.5$ with the empirical EER threshold of the speaker encoder.
For GE2E fine-tuning, each batch contained $N{=}16$ speakers with $M{=}10$ utterances per speaker, and training ran for 30 epochs (AdamW, learning rate $10^{-4}$).

\component{Evaluation Metrics}
We report EER to measure overall discriminability, and FRR at two operating points: FRR@1\% FAR (strict-security setting) and FRR@10\% FAR (more relaxed setting), consistent with PK-MTL~\cite{yang2022}.

\begin{figure}[t]
  \centering
  \includegraphics[width=0.7\linewidth]{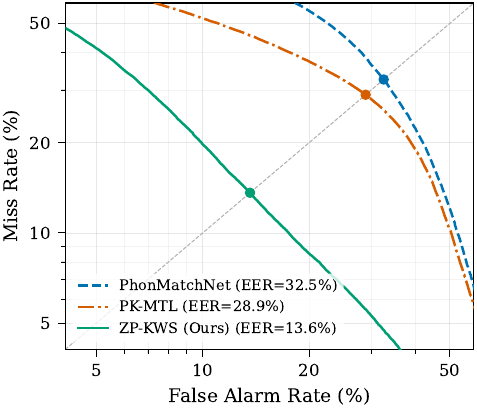}
  \vspace{-10pt}
  \caption{
    Log--log DET curves comparing ZP-KWS against baselines under the stringent TO-KWS operational mode.
  }
  \label{fig:det}
  \vspace{-20pt}
\end{figure}

\subsection{Results and Discussion}

\component{Suppressing Impostor Activations}
Table~\ref{tab:main_results} reports performance across three operational modes.
LibriPhrase Easy and Hard were in-domain evaluations; Google Speech Commands and Qualcomm Keyword Speech served as out-of-domain benchmarks with unseen acoustic conditions and vocabularies.
While all systems performed adequately in conventional C-KWS, the stringent TO-KWS mode clearly exposed the speaker-agnostic vulnerability. In this mode, PhonMatchNet's FRR@1\% rose to 97.00\% on LibriPhrase Easy and 93.12\% on Qualcomm.
PK-MTL partially mitigated this issue through shared-encoder multi-task learning, but FRR@1\% still remained high at 72.79\% and 55.71\%, respectively.
ZP-KWS reduced FRR@1\% to 29.47\% and 33.12\%, corresponding to relative reductions of approximately 60\% and 41\% over PK-MTL, and 70\% and 64\% over PhonMatchNet.
Importantly, these TO-KWS gains did not degrade conventional keyword detection: ZP-KWS achieved the best C-KWS EER on three of four datasets (e.g., 2.38\% vs.\ 3.34\% on LibriPhrase Easy).

\component{Robustness on Hard Minimal Pairs}
On the Hard split---where minimal pairs pushed all C-KWS FRR@1\% above 80\%---FRR@10\% provides a more practical operating view. Under this relaxed point, ZP-KWS remained substantially better (C-KWS 31.69\%, TO-KWS 19.62\%) than PhonMatchNet (TO-KWS 74.94\%), and it also achieved the best TO-KWS EER (13.87\% vs.\ 31.86\%). These results indicate that speaker verification complements keyword matching even under extreme phonetic overlap.

\component{Operating-Point Analysis}
Figure~\ref{fig:det} shows log--log DET curves for the miss rate--false alarm trade-off under TO-KWS.
In the operationally critical low-FAR region (FAR $\le$ 5\%), both baselines exhibited miss rates plateauing above 50\%, reflecting the inability of speaker-agnostic systems to reject impostors at tight thresholds.
ZP-KWS maintained substantially lower miss rates across the entire FAR range, and the gap widened as the operating point became stricter, which is the regime most relevant to always-on edge deployment.
As statistical validation, paired bootstrap tests~\cite{bisani2004} with 1,000 resamples confirmed that these improvements were significant across all four datasets ($p < 0.001$).

\component{Ablation Study}
We ablated three key components in Table~\ref{tab:main_results}.
Overall, the ablations show that GE2E pre-training, calibration, and frame-level phoneme supervision each contribute distinct and complementary gains.

Removing GE2E pre-training caused the largest degradation in target-only performance, with TO-KWS FRR@1\% increasing from 29.47\% to 73.42\% on LibriPhrase Easy and from 33.12\% to 67.44\% on Qualcomm. This pattern is consistent with isolated speaker verification, where GE2E fine-tuning reduced EER from 22.19\% to 8.41\% on LibriPhrase (62\% relative) and from 23.07\% to 12.28\% on GSC (47\% relative), while Qualcomm changed only slightly (6.52\% $\to$ 6.23\%), likely because its longer utterances already produced stable embeddings.

We also found that calibration is necessary to align raw cosine similarities with the semantic probability space: without calibration, TO-KWS EER degraded from 8.16\% to 15.24\% on LibriPhrase Easy and from 8.81\% to 15.00\% on Qualcomm, and the similarity scores clustered near 0.5 with insufficient dynamic range for multiplicative fusion.

In addition, removing frame-level phoneme supervision ($\mathcal{L}_{align}$) degraded C-KWS EER on three of four datasets, with the largest gap on Qualcomm (6.88\% $\to$ 10.81\%), and also incurred severe TO-KWS penalties (e.g., Qualcomm FRR@1\%: 33.12\% $\to$ 43.64\%); this suggests that on tasks with low inter-class phonetic overlap, $\mathcal{L}_{align}$ can over-regularize by enforcing fine-grained phonetic structure when coarser cues may suffice.

Finally, element-wise minimum fusion ($p_{final} = \min(p_{utt}, p_{spk})$) produced results comparable to multiplication, indicating that the primary gain comes from AND-style gating rather than the specific fusion operator.

\component{Feasibility for Edge Deployment}
ZP-KWS uses approximately 1.55\,M parameters in total (0.65\,M for the KWS branch and 0.90\,M for EfficientTDNN-Small).
This footprint is suitable for on-device deployment, and runtime overhead remains low: the speaker encoder for enrollment is frozen and computed once per user, while each query requires one cosine-similarity calculation followed by scalar multiplication for late fusion.

\section{Conclusion and Future Work}
This work demonstrates that user-defined keyword spotting can incorporate biometric security without sacrificing zero-shot keyword generalization.
Our central finding is that text-independent speaker verification becomes practical for short-utterance UD-KWS when the architecture is functionally decoupled: i) a GE2E-pretrained compact speaker encoder stabilizes sub-second speaker embeddings, ii) frame-level phoneme supervision strengthens keyword-relevant acoustic structure, and iii) multiplicative late fusion enforces veto-style identity gating at inference.
Under this design, a single trained model can switch between conventional and strict speaker-gated operation without retraining.
Experimentally, ZP-KWS reduces TO-KWS FRR@1\% FAR by up to 60\% relative to the strongest baseline while staying within a 1.55\,M-parameter budget, supporting feasibility for always-on edge scenarios.
In future work, we will focus on confidence calibration under noisy and mismatched conditions to further improve robustness in the dual zero-shot setting.

\newpage

\section{Generative AI Use Disclosure}

Claude Opus 4.5 was used only for limited coding assistance (e.g., debugging suggestions). Gemini 3 Pro, ChatGPT, and Prism were used only for language refinement and editorial polishing. All study design, implementation, experiments, and final scientific decisions were performed and verified by the authors.

\section{Acknowledgment}

This work was supported in part by Realtek Semiconductor Corporation under Grant Numbers 113KK01103 and 114KK01005. Any findings and implications in the paper do not necessarily reflect those of the sponsors.

\bibliographystyle{IEEEtran}
\begingroup
\bibliography{references}
\endgroup

\end{document}